%% file: main.tex
\documentclass[twocolumn,letterpaper, 9.5pt]{article}

\DeclareMathSizes{10}{10.5}{6.5}{6.5}
\usepackage{anysize}
\usepackage{fancyhdr}
 
\marginsize{1.75cm}{1.75cm}{0.5cm}{0.5cm}
 \usepackage[inkscapelatex=false]{svg}

\usepackage[symbol]{footmisc}

\newlength{\wdth}

\input{Packages}

\usepackage{xcolor}

\begin{document}

\title{
	\vspace{-1.5cm}\rule{\linewidth}{4pt}\vspace{0.3cm} \Large \textbf{
    Leveraging Metrologically Useful States in Quantum Reservoir Networks \\ for Latent Space PDE Predictions
	}\\ \rule{\linewidth}{1.5pt}}
	\author{\stepcounter{footnote}Erik L. Connerty\textsuperscript{1}\thanks{Corresponding author. Email: \href{erikc@cec.sc.edu}{erikc@cec.sc.edu}}, Margarite   LaBorde\textsuperscript{2,3}, 
     and Ethan N. Evans\textsuperscript{4} \\
    \vspace{-.0cm}
	\small{\textsuperscript{1}University of South Carolina - Columbia, Columbia, SC, USA} \vspace{-0.0cm} \\
    \small{\textsuperscript{2}Naval Surface Warfare Center Panama City Division, Panama City, FL, USA} \vspace{-0.0cm} \\
    \small{\textsuperscript{3}The Johns Hopkins Applied Physics Laboratory, Laurel, MD, USA} \vspace{-0.0cm} \\
    \small{\textsuperscript{4}Qodex Quantum, Chicago, IL, USA} \vspace{-0.0cm}
	}
        \date{\vspace{-.5cm}}
	\maketitle

\begin{abstract}
    Interest in using quantum computers for the purpose of predicting chaotic partial differential equations (PDEs) has been growing with the advent of newer low-error quantum computers and robust simulation tools. In this paper, we present a method that utilizes a quantum reservoir network (QRN) to predict latent space representations of the high-dimensional chaotic 1-D Kuramoto-Sivashinksy (KS) system. This hybrid approach takes advantage of advancements in classical machine learning (ML) through the use of a classical autoencoder as well as techniques from quantum metrology through the use of a unitary that creates metrologically-useful states. Through rigorous simulation and analysis, we show that the proposed method outperforms alternative QRN implementations without this metrologically-useful state preparation, and also show better performance than classical echo-state networks when weight regularization is not used. Finally, we bring to light potential issues that can arise when using autoencoders within QRC pipelines.
    
\end{abstract}

\section{Introduction}

Quantum computers have shown great promise in solving traditionally intractable problems, or improving upon known classical limits \cite{Shor_1997, grover10.1145/237814.237866}. However, the use of quantum computers for the purpose of general machine learning tasks remains a niche topic of interest with many findings failing to be replicated on hardware due to various challenges and constraints. With this in mind, we introduce a method for PDE prediction on quantum computers which uses a dynamical quantum reservoir (aptly called a quantum reservoir network (QRN)) coupled with a classical autoencoder to perform latent space prediction of the chaotic Kuramoto-Sivashinksy (KS) system. This QRN builds upon the design in \cite{Connerty2026} by including the addition of input-dependent rotations on all two-qubit gates, a  weight-initialization scheme that guarantees the generation of \textit{genuine multipartite entanglement}, and the addition of \textit{metrologically-useful} state preparation and measurement procedures. These changes give access to an even larger amount of expressivity in the input-to-output (I/O) mapping and allow us to successfully predict the 128-dimensional 1-D KS system with high accuracy. In addition, this paper investigates the relationship between different unitary designs, the number of shots required to converge on an optimal solution to the PDE prediction problem, as well as the quantum Fisher information (QFI) across configurations.

\paragraph{Related Works:}
Reservoir networks have historically been used to predict the dynamics of chaotic time-series data including PDEs \cite{jaeger2001}. The field of quantum reservoir computing (QRC) has also been quite promising in this regard, but currently, most works are still focusing on lower-dimensional systems \cite{Connerty2026, krisnanda_experimental_2025, kornjača2024largescalequantumreservoirlearning, Hu2024,ahmed10.1098/rspa.2025.0550, Murauer_2025,feedback_PRXQuantum.5.040325,fullmeas_PhysRevResearch.6.013051,pfefferPhysRevResearch.4.033176,Mujal2023,yasuda2023quantumreservoircomputingrepeated,monomi2025feedbackenhancedquantumreservoircomputing,feed_Zhu2025,feed_Paparelle2026-qj}. Much recent work focuses on solving PDEs using quantum variational algorithms (QVA) \cite{Ali2023Poisson,Guseynv2023heat,LIU2024Stokes,Sarma2024nonlinearPDE,Hunout2025Lagrange,Choi2025Inhomogeneous} which often involve simulating Hamiltonian evolutions \cite{Li2023Poisson,alipanah2025diffusion,SONG2025NavierStokes}, a process which is known to be resource intensive if the Hamiltonian is sufficiently non-local and not sparse. Jumping to larger-dimensional systems requires innovative approaches to data-embedding and reconstruction, which have largely gone unexplored in QRC. For variational circuits, works such as \cite{chen2025quantumrecurrentneuralnetworks,takagi2025timeseriesforecastingnonlinearhighdimensional} have explored training quantum circuits for time-series prediction while using a classical autoencoder, but in contrast to recurrent QRN designs \cite{Hu2024,Connerty2026}, these circuits do not contain true memory in the quantum sense and have not demonstrated hardware implementations beyond the $T_1$ and $T_2$ times of a superconducting QPU. Our implementation, as well as the aforementioned works, utilize this classical autoencoder to help with dimensionality reduction which can be an important step in PDE prediction tasks.


\section{Method}\label{sec:methods}

Predicting chaotic PDEs such as the KS system described in \cite{kuramoto10.1143/PTPS.64.346, SIVASHINSKY19771177} on quantum computers requires novel approaches due to the challenges that arise with high-dimension input data and noisy intermediate-scale quantum computing (NISQ) hardware, as well as certain constraints when creating QRCs with persistent memory. In particular, difficulty arises when trying to embed data using angle encoding if the dimensionality is too high, as the number of qubits that are easy to simulate and test on current hardware is typically low. Naïvely, one may question why the popular \textit{amplitude embedding} is not used in this context, as the scaling is logarithmic in the number of qubits; however, 
while previous embedding and readout schemes in QRNs can be shown to have persistent memory, the same schemes using an amplitude embedding would not. This is due to the amplitude embedding's inability to facilitate continuous quantum state evolution. With this in mind, we modify the angle embedding from \cite{Connerty2026} to be more expressive, remove sparsity from the two-qubit entangling gates, and utilize a GHZ unitary to create what we term a \textit{metrologically-useful} state. To reduce the dimensionality of the original PDE, we utilize a classical autoencoder to make the embedding more tractable for low-qubit counts. These methods are inspired by a range of both quantum and classical methods in machine learning and metrology as detailed in \cite{metrologygezaPhysRevA.85.022322,metrologygezaPhysRevLett.125.020402, latentLI2025113705, latentKontolati2024}.

\subsection{Quantum Reservoir Network}\label{sec:QRN}

\begin{figure*}[h!]
    \centering
    \resizebox{\textwidth}{!}{\input{figures/nn_GHZ}}
    \caption{The GHZ QRN Circuit. As before, the inner block named the ``Reuploading Block'' is repeated $n_c = 3$ times in all experiments, as this reuploadng procedure has been shown to improve performance. The outer ``Recurrent Block'' is repeated for every time-step of the input data. A ``GHZ'' unitary is performed at the start of every ``Recurrent Block'' and serves to create an optimal reservoir-like state. The ``CRZ'' operation at the end of each ``Reuploading Block'' has been modified to have only adjacent connections, allowing for no SWAP operations when mapping to an IBM Heron R2 QPU. At the end of each ``Recurrent Block'', a measurement is taken on half of the qubits, and those qubits are reset to $\ket{0}$, creating a weak-measurement scheme that retains memory.}
    \label{fig:qesn_circuit_GHZ}
\end{figure*}
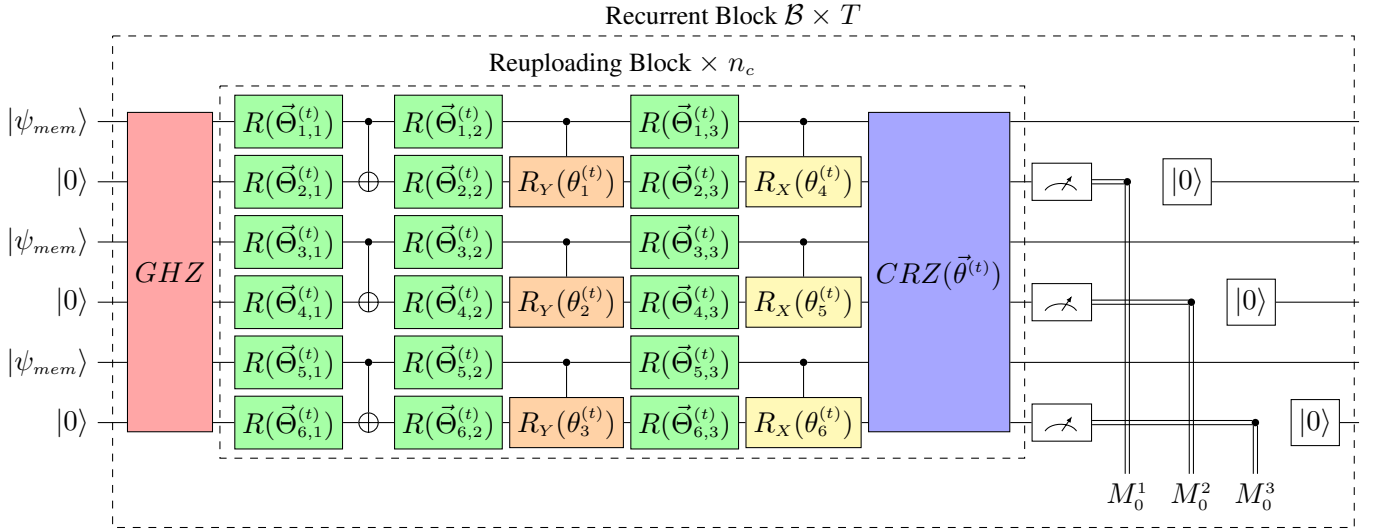

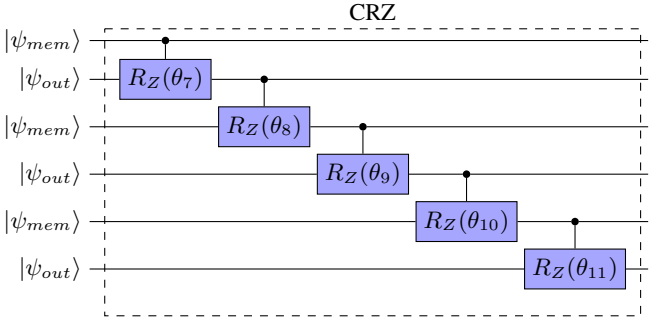
\begin{figure}[h!]
    \centering
    \input{figures/CRZ}
    \caption{Circuit depiction of the cascaded CRZ gate used in the QRN.}
    \label{fig:CRZ}
\end{figure}

The QRN architecture used here is depicted in Figure~\ref{fig:qesn_circuit_GHZ} and features an expressive input-embedding scheme as well as metrology-inspired state-preparation. For the embeddings, unique weights are generated for every single rotation gate $R(\Theta)$ in the circuit, and all two-qubit gates other than CNOT receive both an input-dependent rotation as well as a bias rotation. Additionally, the ``Reuploading block'' is constructed to guarantee multipartite entanglement, and a GHZ-like unitary is performed at each timestep to mimic metrology-based state-creation procedures. Finally, all non-adjacent operations are removed from the $CRZ$ operations at the end of the ``Reuploading Layer'' to create a new unitary optimized for IBM Heron QPUs (Quantum Processing Units) detailed in Figure \ref{fig:CRZ}.


\paragraph{Data Mapping}  
Real-valued time-series data is mapped into rotation angles by a tensor contraction that is performed on the input data and the circuit's generated weights. Let the input $x_i(t)$ be the $i$th value in the input vector at time $t$ where $t \in \set{1,2,\dots,N}$ and $N$ is the length of the time-series. The index $i$ is given as $i \in \{1,...,d\}$ where $d$ is the dimension of the input vector. We encode the classical data at each time $t$ onto $n_q$ qubits via three rotation gates per qubit and three Euler angles per rotation gate indexed by $j,k,$ and $l$, respectively. The input weight tensor and the bias weight tensor are given as $W^{in} \in \mathbb{R}^{d \times n_q \times 3 \times 3}$ and $W^{bias} \in \mathbb{R}^{n_q \times 3 \times 3}$. This gives the output rotation angles $\Theta \in \mathbb{R}^{n_q \times 3 \times 3}$ as a dot product between the input $x(t)$ and the weight tensor $W^{in}$ added to the bias tensor $W^{bias}$ as
\begin{align}\label{eq:input-embedding}
\Theta^{(t)}_{j,k,l} = \sum_{i=1}^{d} x_{i}(t) \, W^{in}_{i,j,k,l} + W^{bias}_{j,k,l},
\end{align}
Thus, the entry $\Theta_{1,2,3}$ refers to the third Euler angle on the second rotation gate for the first qubit in the circuit. Two-qubit gate rotations are defined similarly, with the exception that their weight tensors do not include $3$ Euler angles, and instead only have one rotation angle.

\paragraph{Angle Embedding:}
The computed angles from Eq.~\eqref{eq:input-embedding} then construct the gate
\begin{equation}
R_{jk}^{(t)}(\Theta_1,\Theta_2,\Theta_3)
=
R_z(\Theta_1)R_x(\Theta_2)R_z(\Theta_3),
\label{eq:rotation}
\end{equation}

where $R_x$ and $R_z$ are arbitrary single-qubit rotation gates, $\Theta^{(t)}$ is an input-dependent Euler angle rotation matrix, and $R_{jk}^{(t)}$ is the $k$th rotation gate on the $j$th memory qubit at time-step $t$. Data is also mapped to all two-qubit gates except CNOT using a single rotation angle computed similarly from Eq.~\eqref{eq:input-embedding}, which will be denoted as
    \begin{equation}
    CU(\theta),
    \end{equation}
    and references the controlled rotations in Figure~\ref{fig:qesn_circuit_GHZ}.

\paragraph{Weight Initialization:}
We additionally define the tensors
$W^{hidden} \in \mathbb{R}^{\frac{n_q}{2} \times 2}$,
$W^{in\_hidden} \in \mathbb{R}^{d \times \frac{n_q}{2} \times 2}$,
$W^{entangle} \in \mathbb{R}^{n_q-1}$, and
$W^{in\_entangle} \in \mathbb{R}^{d \times n_q}$,
which are all scaled by the constant $C = d  n_c$,
where $n_c$ is the number of repeats of the ``Reuploading Block'' detailed in Figure \ref{fig:qesn_circuit_GHZ} and $d$ is the dimensionality of the input vector as described before. The weight scaling for the input weight tensor is sampled from a uniform distribution, $W^{in} \sim U(\frac{-\pi}{C}, \frac{\pi}{C})$.
Bias weights for each gate are sampled similarly as $W^{bias}, W^{hidden}, W^{entangle} \sim U(\frac{-\pi}{2 n_c}, \frac{\pi}{2 n_c})$
Finally, input weights for two-qubits gates are sampled as $W^{in\_hidden}, W^{in\_entangle} \sim U(0,\frac{\pi}{C})$.
This controlled weight scheme normalized by the dimensionality of the data and number of re-euploading blocks ensures individual data points do not perturb the network too greatly.

\paragraph{GHZ Unitary:}

Previous work on QRNs used an initial input state of $\ket{0}^{\otimes n}$ \cite{Connerty2026,Hu2024}.
Although natural from an implementation perspective, this initialization introduces
a strong structural bias: before any nontrivial evolution, all probability weight is
concentrated on a single computational-basis state. Moreover, for shallow circuits
or for parameter updates close to identity, the evolved state can remain
strongly localized near the all-zero configuration. As a result, the induced
measurement distribution may be highly peaked, with many basis states having
very small probabilities.

This concentration is not necessarily problematic in metrological protocols where
the relevant signal may be encoded in the probability of a particular outcome,
often the all-zero outcome. In contrast, the reservoir network considered here
uses the measured output probabilities over the computational basis to construct
a feature vector. In this setting, a highly concentrated distribution can be
undesirable: basis states with very small probabilities are estimated with larger
relative sampling noise and may contribute less reliably to the learned feature
representation. We therefore seek initializations that distribute amplitude more
symmetrically across the computational basis while still retaining sensitivity to
parameter-dependent evolution.

An example of such a symmetric state is given by the GHZ state
\begin{equation}
\ket{GHZ} = \frac{\ket{0}^{\otimes n} + \ket{1}^{\otimes n}}{\sqrt{2}}
\end{equation}
This state generates an equal superposition over two fully polarized states, allowing for further perturbations to generate output probabilities without biasing towards one extrema over the other. We hypothesize that this symmetric state reduces the number of measurements needed to fully reconstruct our probability density function. 

We also motivate this choice of initial state through a metrology-based perspective. GHZ states are used in quantum metrology to achieve Heisenberg-scaled sensitivity \cite{Leibfried2004spectroscopy,Tóth_2014} for parameter estimation tasks. Separable states, such as the fully polarized $\ket{0}^{\otimes n}$ state, are known to be shot-noise limited in their sensitivity, such that the precision of any measurement using these states scales as $\Delta\theta = 1/\sqrt{N}$ with the number of qubits. In comparison, states with Heisenberg-scaled sensitivity boast an enhanced precision of $\Delta\theta = 1/{N}$. While entanglement is necessary to achieve sub-shot-noise scaling \cite{Pezze2009entanglement,Giovannetti2006Metrology}, many entangled states still do not achieve non-classical sensitivity \cite{Hyllus2010Notall}. Thus, the GHZ state is a particularly useful state from a metrological perspective. As quantum machine learning also involves extracting information from input data, it is natural to expect that similar gains can be observed in this context.

\begin{figure}[t]
    \centering
    \input{figures/ghz}
    \caption{State creation circuit for the GHZ state, which is depicted in the QRN circuit as a multiqubit GHZ.}
    \label{fig:GHZ}
\end{figure}
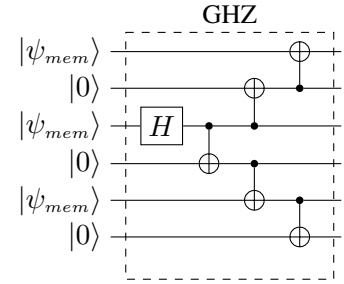

To fully take advantage of the QRN's recurrent nature and memory, we apply a GHZ state preparation unitary at each timestep, even though its input state is constantly evolving. This repeated application creates a bias towards GHZ-like states and has the desirable effect of increasing the circuit's performance on the latent space prediction task detailed in Section~\ref{sec:results}. The optimized unitary for the $GHZ$ block is shown in Figure~\ref{fig:GHZ}.

\subsection{Convolutional Autoencoder}
\begin{figure*}[h!]
  \centering
  \includegraphics[width=\textwidth,keepaspectratio]{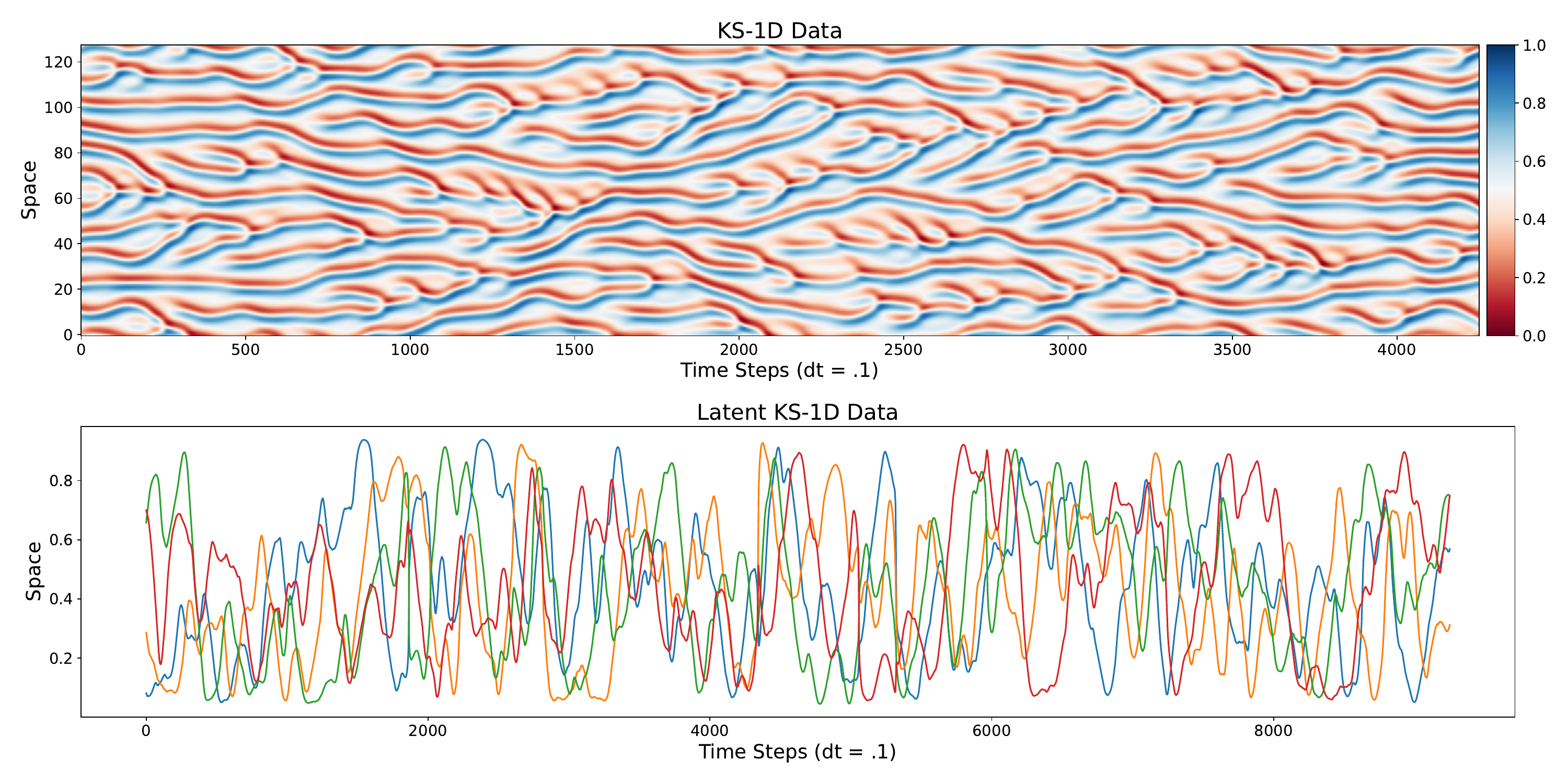}
  \caption{1-D Kuramoto Sivashinsky data. The top data is the original data with $n = 128$ degrees of freedom, the bottom is the latent space representation with only $n = 4$.}
  \label{fig:ks-1d}
\end{figure*}

Dimensionality reduction is an important part of the PDE prediction architecture. In order to handle the 128-D input of the KS system, a standard residual-net convolutional autoencoder with $n\_params = 1,085,925$ \cite{resnet7780459} was trained using 9,250 data points sampled from the simulated KS system. The activation function $\sigma$ at both the bottleneck layer and output layer was chosen to be
\begin{equation}
    \sigma(x) = \frac{1}{1+e^{-x}}
\end{equation}
to guarantee that all latent space variables were contained in the interval $[0,1]$. This is done to ensure that input-embeddings remain consistent under the QRN's weight initialization scheme. Both the training data and latent space variables derived from \cite{ksjax} for the autoencoder are shown in Figure \ref{fig:ks-1d}.

\subsection{PDE Prediction Architecture}
To make the task of embedding high-dimensional data into the QRN circuit more tractable with lower qubit counts, the convolutional autoencoder described in Section \ref{sec:methods} reduced the 128 degree-of-freedom (DOF) PDE into a 4 DOF system as shown in Figure \ref{fig:ks-1d}. It is speculated that having access to a larger amount of qubits would have made this process unnecessary, but in the interest of performing experiments on current hardware and in simulation, the autoencoder is considered an instrumental component of the PDE prediction architecture.

Overall, the QRN is used to predict latent space representations of the KS system that have been harvested from the autoencoder's bottleneck layer. These values are fed into the QRN as detailed in Section \ref{sec:methods}. At each time-step of the circuit's evolution, partial measurements are taken on the \textit{Readout} qubits, which are interspersed with the \textit{Memory} qubits as shown in Figure \ref{fig:qesn_circuit_GHZ}. After measurements for all $N_t$ time-steps are recorded, a linear regression is performed to fit the measurements to the $(t+1)$ latent space representation of the KS system. These predictions are then fed back through the decoder and projected back into the original $n = 128$ DOF PDE. 


\section{Simulation Results}\label{sec:results}
Simulated results are gathered using a $n_q = 16$ qubit circuit with the Aer simulator on a DGX-A6000 NVIDIA cluster. Measurements are collected in 10,000 shot increments up to 960,000 total shots to fully recover an accurate distribution for an entire circuit run and to show the scaling for different circuits detailed in Section \ref{subsec:scaling}. The total time-series length used was $N_t = 5000$, where $n_{train} = 3000$ were the number of data points used for training and $n_{test} = 2000$ were used for test validation. A washout length of $n_{wash} = 20$ was discarded from the training data before fitting, as is typical in reservoir computing literature, giving the system time to ``washout'' the effect of initial conditions.

\subsection{PDE Predictions}\label{subsec:pde_predictions}
\begin{figure}[h!]
  \centering
  \includegraphics[width=\columnwidth,keepaspectratio]{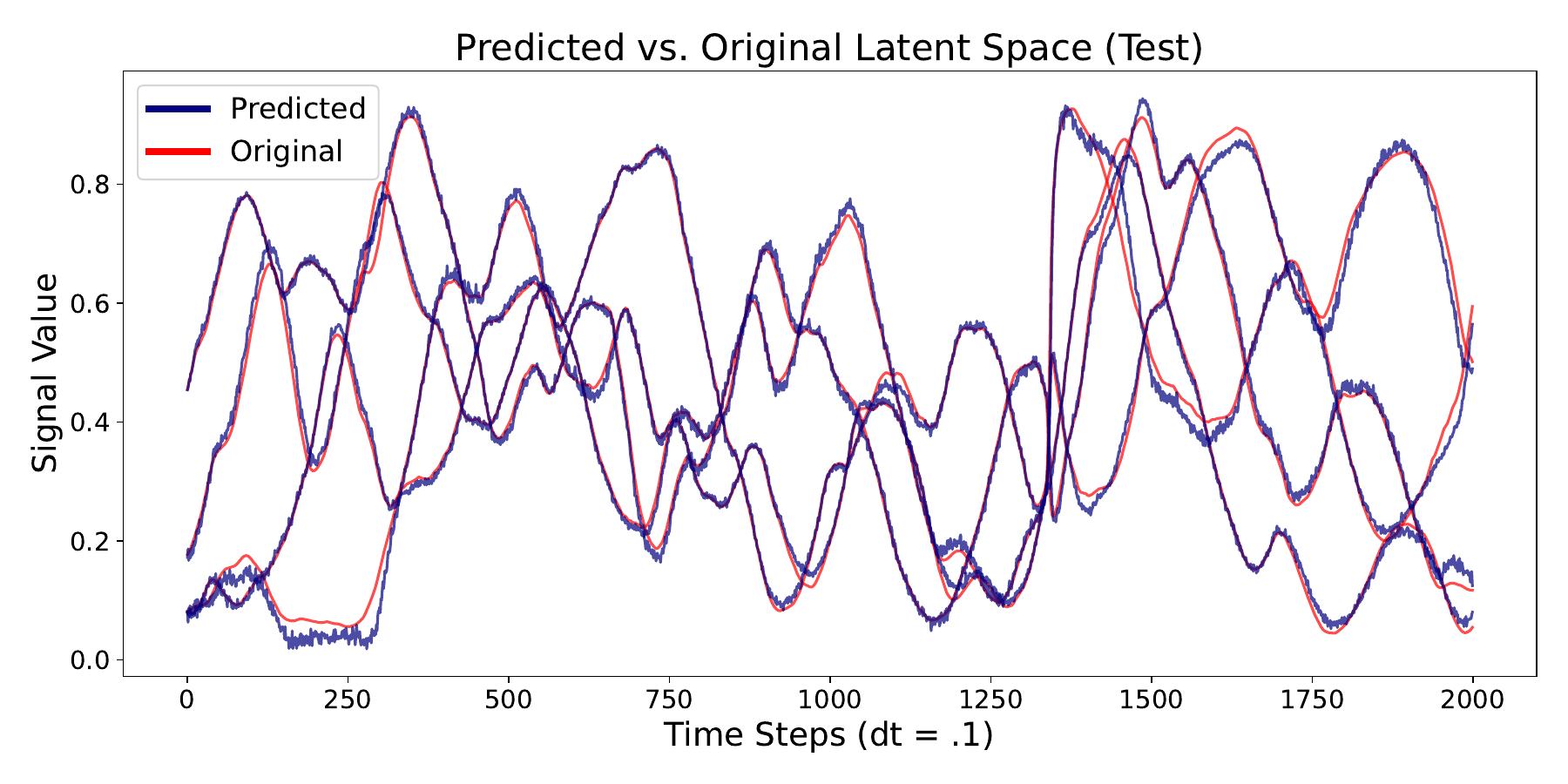}
  \caption{Latent space test data performance of the QRN with GHZ block.}
  \label{fig:latent_test_comparison}
\end{figure}

\begin{figure*}[h!]
  \centering
  \includegraphics[width=\textwidth,keepaspectratio]{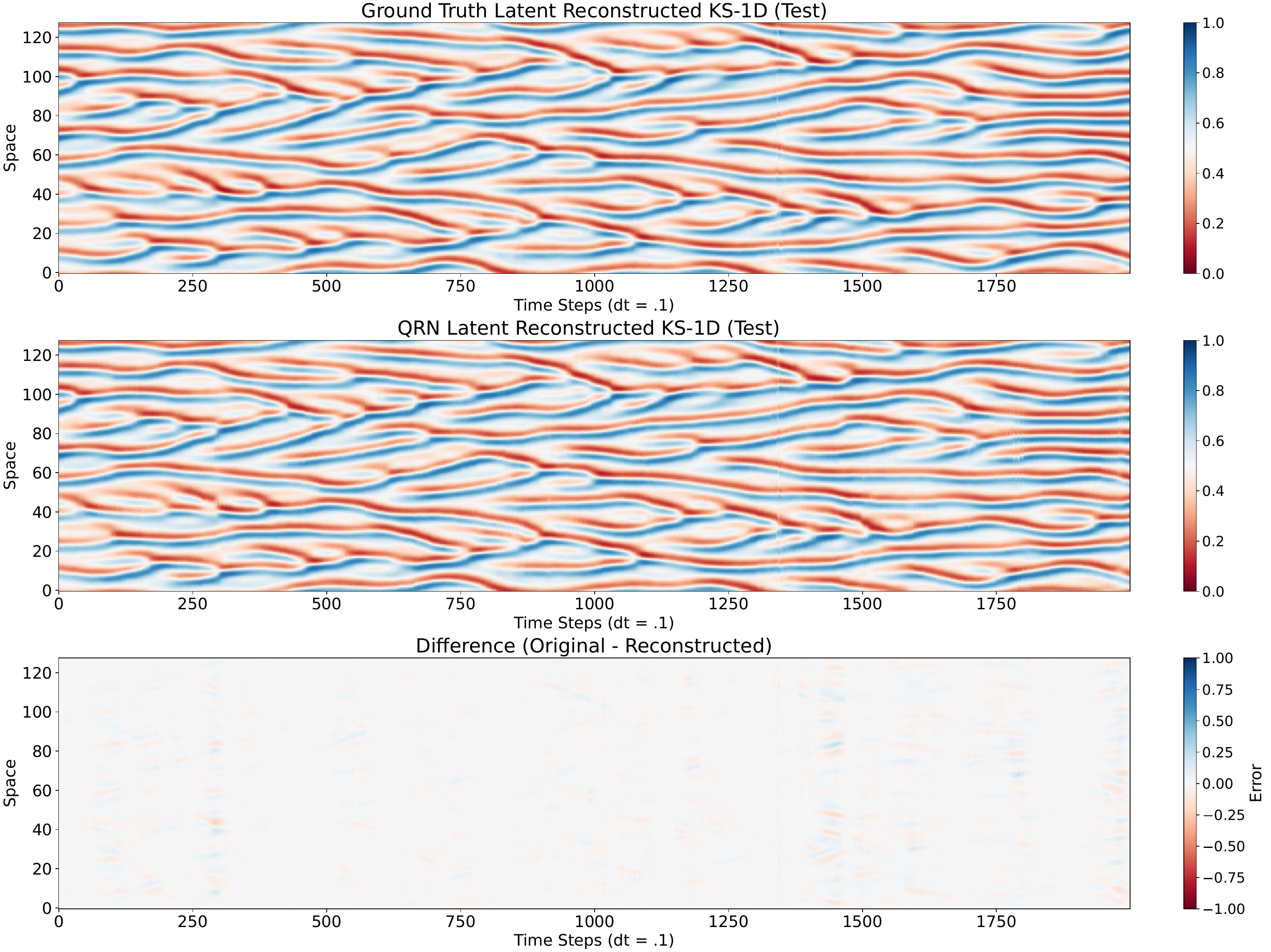}
  \caption{Decoded latent space representation of the KS-1D system compared to the ground truth decoded latents for $(t+1)$ predictions. The top figure denotes the decoded ground truth latent. The middle figure shows the output after being passed through the QRN for time-series prediction and then through the decoder for reconstruction. The bottom figure shows the residual between the reconstructed ground truth and the reconstructed prediction. Notably, the autoencoder itself introduces some small errors to the reconstruction, but the results of the QRN are close to the original with the $(t+1)$ prediction task.}
  \label{fig:ks_test_decode}
\end{figure*}

We report that the QRN architecture detailed in Section \ref{sec:methods} exhibits strong performance on the $(t+1)$ prediction task. In comparison to the previous method specified in \cite{Connerty2026} and other versions of the circuit that did not have the added GHZ unitary, the performance is an order-of-magnitude better. These results are shown in Figure \ref{fig:latent_test_comparison} on the test set latent space KS system, where a root mean square error (RMSE) of $.0162$ is achieved. This result gives a low enough error for the decoder to properly reconstruct the KS PDE with high accuracy, as shown in Figure~\ref{fig:ks_test_decode}. 
The decoded output is nearly on par with the original KS data in its original form before being passed through the autoencoder or QRN pipeline. It should be understood that the autoencoder itself introduces a small amount of error in the reconstruction process, even with ground truth data, and that for the loss to be so minimal, the latent space predictions must be of very high accuracy.

\subsection{Metrologically Useful States and Shot-Scaling}\label{subsec:scaling}
To better understand the differences in the various versions of the QRN and the effect of the GHZ unitary, a shot-scaling analysis was performed on 4 versions of the circuit: QRN with GHZ state preparation, QRN without GHZ, Sparse QRN, and a QRN with random state preparation. The Sparse QRN is the same circuit described in \cite{Connerty2026}, which used sparsity to reduce circuit depth, potential gate errors, and generate random unitaries. However, the Sparse QRN was found to be unsuitable for this more difficult task because of the separability of the quantum states and less expressive data embeddings. 

As discussed in Section~\ref{sec:QRN}, the QRN with GHZ state preparation was employed instead to leverage the entanglement and Heisenberg-limited sensitivity of the GHZ state. This circuit features additional changes, such as the previously discussed updated angle embedding and CRZ layer. The QRN without GHZ is the same general circuit design as the QRN with GHZ, minus the GHZ unitary block at the start of each ``Recurrent Block'' to account for the performance improvement provided by these architectural changes.

For the QRN with Random State Prep., a random state preparation circuit $U$ was implemented at the beginning of each ``Recurrent Block'' to compare directly to the performance of the GHZ unitary. For these runs, multiple different unitaries were generated, and the performance was averaged together, with each unitary being held constant for an entire run of testing and training. The averaged behavior of these systems serves as a comparison to the GHZ procedure, as random unitaries should not be able to achieve the same performance if indeed the GHZ state's metrological sensitivity is responsible for the improvements we observe as \textit{metrologically useful} states which are known to be sparse \cite{Hyllus2010Notall}. Ideally, these $U$ would be Haar-random unitaries across all input qubits; however, sampling from the Haar measure is computationally expensive. Unitary $t$-designs can be employed in polynomial depth \cite{Arnaud2008random,Haferkamp2022randomquantum,Brandão2016} for high-accuracy approximations of Haar-random unitaries, but these approaches still result in large-depth circuits under the QRN construction. For our purposes, we use an efficient random circuit approximation \cite{Schuster2025random} as a proof-of-concept comparison point. 

\begin{figure}[h!]
  \centering
  \includegraphics[width=\columnwidth,keepaspectratio]{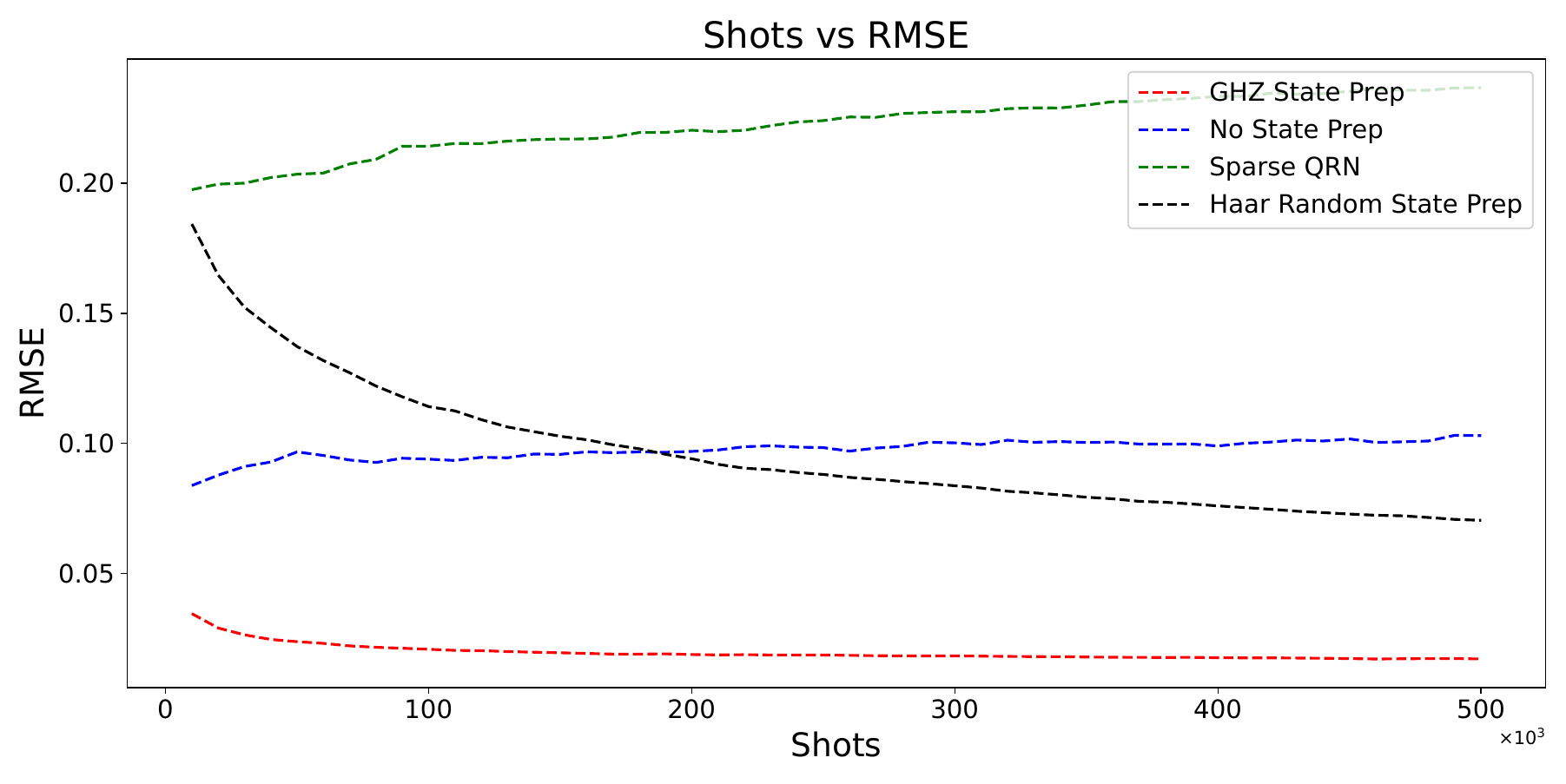}
  \caption{An analysis with the 4 different versions of the QRN with $n_{shots}$ ranging from $10,000 - 500,000$.}
  \label{fig:shot_scaling_analysis}
\end{figure}

The shot-scaling analysis shows how quickly the QRN can reach equilibrium states on the output probability densities and gives a rough upper and lower bound of performance for each version of the circuit in terms of RMSE. Figure~\ref{fig:shot_scaling_analysis} shows an analysis with $n_{shots}$ varying from $10,000 \-- 500,000$, as performance does not seem to improve much beyond this number for any circuit except the QRN with GHZ State Prep. We also note that the performance of the QRN with GHZ State Prep is an order of magnitude better than the Sparse QRN, which was first described in \cite{Connerty2026} and the first QRC circuit to run 100 times over the $T_1$ and $T_2$ times of an IBM QPU and successfully perform accurate time-series prediction. Additional observations and results from these experiments are detailed in Supplementary Note~\ref{SN:Data}.


\subsubsection{Quantum Fisher Information with QRNs}
\begin{figure}[h!]
  \centering
  \includegraphics[width=\columnwidth,keepaspectratio]{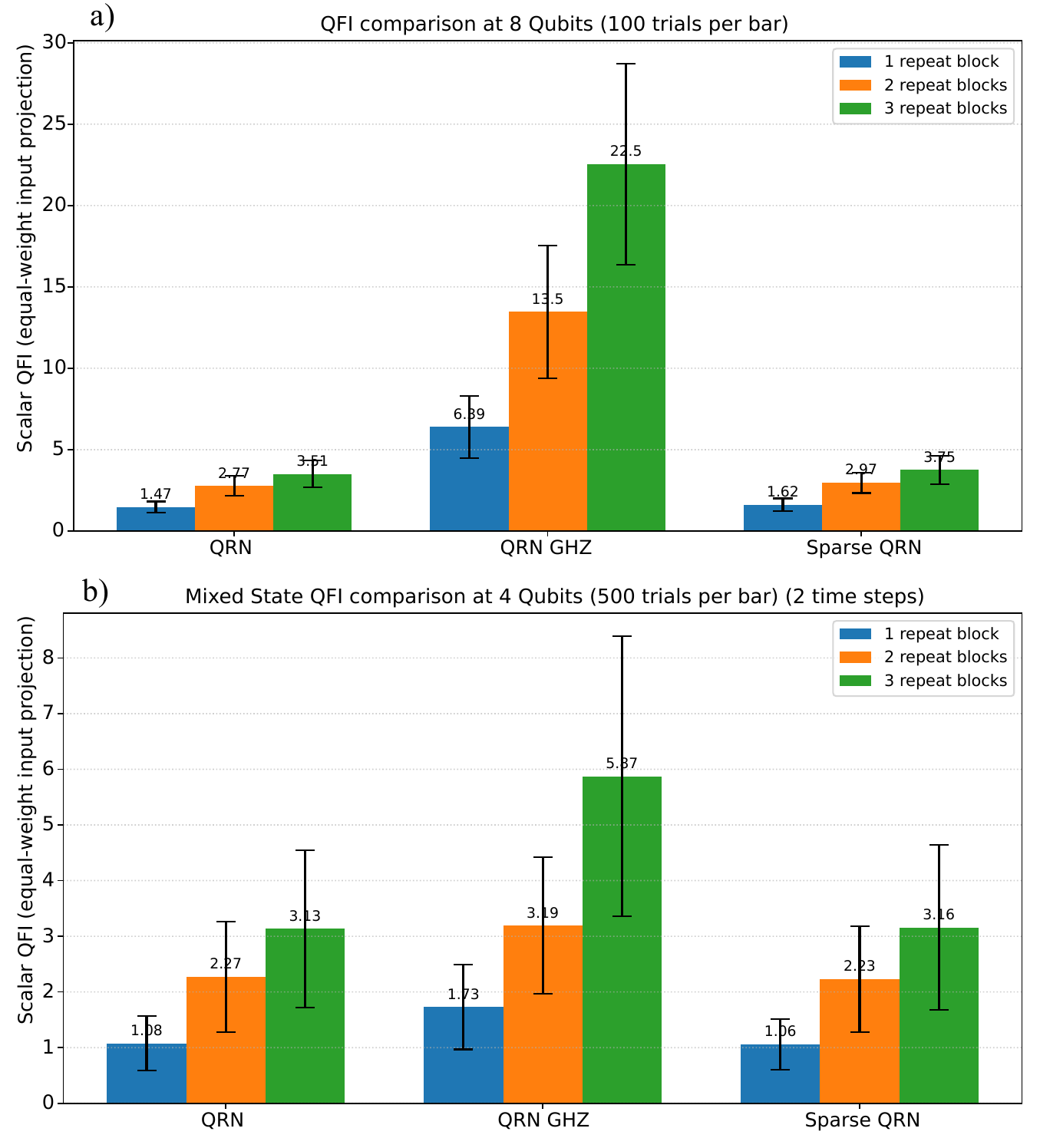}
  \caption{Combined QFI comparison. a) The pure state QFI comparison for $n_{qubits} = 8$ shows a marked increase for QFI in the QRN GHZ case, and also shows that added reuploading blocks increases QFI. b) The mixed state result with $n_{qubits} = 4$ shows similar trends, but with higher variance. The pure state results are collected over 100 trials, while the mixed state results used 500 trials.}
  \label{fig:qfi_combined}
\end{figure}

We compare quantum fisher information (QFI) for two configurations of the QRN to determine whether the QRN with GHZ increases QFI. In Figure~\ref{fig:qfi_combined} (a) we show that the QRN with GHZ has improved QFI by taking the average QFI matrix entry over all parameters when looking at the pure state before non-unitary measurement and reset. This QFI matrix is given as
\begin{equation}
    F_Q(\theta) = 4 \left[ \langle \partial_\theta \psi_\theta | \partial_\theta \psi_\theta \rangle - | \langle \psi_\theta | \partial_\theta \psi_\theta \rangle |^2 \right],
\end{equation}
where every rotation gate is parameterized.
For the case of mixed-state QFI that occurs after non-unitary measurement and reset, we calculate the QFI as
\begin{equation}
    F_Q^{ij}[\rho(\boldsymbol{\theta})] = 2 \sum_{k,l} \frac{\text{Re}(\langle k | \partial_i \rho | l \rangle \langle l | \partial_j \rho | k \rangle)}{\lambda_k + \lambda_l},
\end{equation}
and take the average of all QFI matrix entries as before. This is shown in Figure~\ref{fig:qfi_combined} (b). We note that the high-variance of the mixed-state QFI experiment over different seeds makes it inconclusive.

\subsection{Comparison to Classical Echo-state Network}
\begin{figure}[h!]
  \centering
  \includegraphics[width=\columnwidth,keepaspectratio]{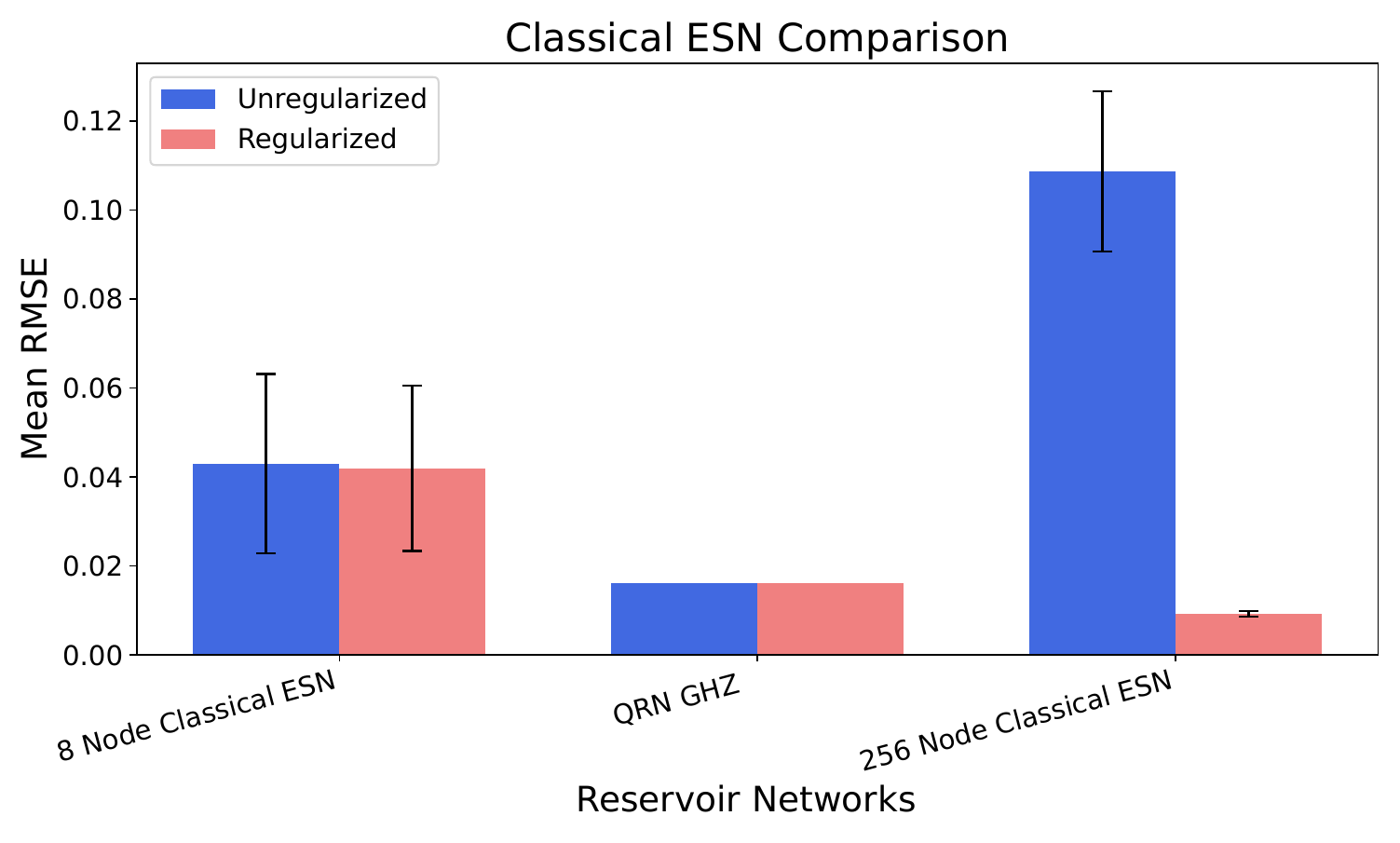}
  \caption{A comparison showing the differing performance between an 8 and 256 node classical ESN vs the QRN. Ridge regularization parameters of $\alpha = .1$ were used for the classical ESNs in the ``Regularized'' cases. The QRN used $\alpha = 0$ as it did not perform any better with regularization. A sample size of $n = 10$ was used for the classical ESNs. For the QRN, computational complexity was prohibitively large and only allowed for the plotting of the best run.}
  \label{fig:classical_comparison}
\end{figure}

A quick comparison was made to a standard implementation of a classical echo-state network (ESN) with varying numbers of reservoir nodes of $n = 8$ and $n = 256$. These values of classical reservoir nodes were chosen specifically because they had an equal number of readout nodes as the QRN had readout qubits. Notably, for the $n_q = 16$ QRN circuit, only 8 of the qubits are measured, making it a fair assumption to compare it to the $n = 8$ reservoir node ESN; however, the authors also acknowledge that the inherent quantum mechanical properties of qubits allow for 8 \textit{Readout} qubits to produce 256 output features. Thus, an $n = 256$ reservoir node ESN was included. The parameters for the ESN are shown in Table~\ref{tab:esn_parameters}, and the results for the experiments are shown in Figure \ref{fig:classical_comparison}. Interestingly, we find that the QRN performs best without any regularization whatsoever, using an unbiased linear regression model, whereas the classical models require regularization to achieve their best test performance. In order to prevent the $n = 256$ node ESN from overfitting, ridge regression \cite{ridgef2890069-85a1-3da4-a804-1978ce89c3d4} with $\alpha = .1$ was used. This was also used for the $n = 8$ reservoir node ESN. In the comparison chart with the regularized results, the QRN was optimized with $\alpha = 0$, as this produced the best results. While the absolute lowest error was achieved by the $n = 256$ node ESN, the authors still believe the $n = 8$ node comparison was the fairest comparison, and we highlight that it appears the QRN generalizes better to the test data without any need for regularization.

\subsection{Scaling for Different Time-Horizons}
\begin{figure}[h!]
  \centering
  \includegraphics[width=\columnwidth,keepaspectratio]{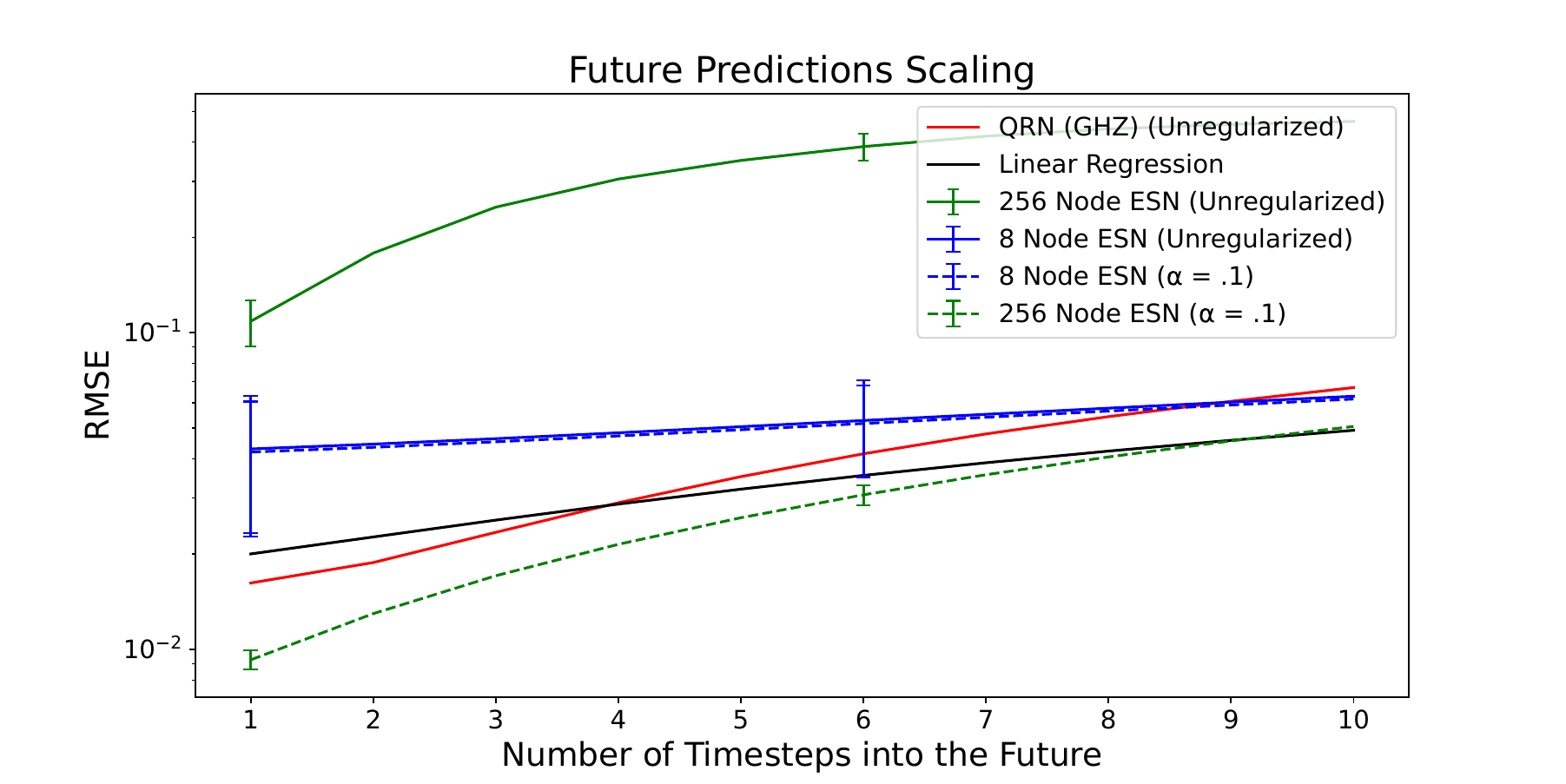}
  \caption{Prediction using varying time-horizons from $(t+1)$ up to $(t+10)$. This experiment with the QRN did not require any sort of regularization to achieve high generalization capabilities on future time-steps and unseen data, while the classical ESN required Ridge regression with $\alpha = .1$. The best performing model being linear regression highights the difficulties that can arise when using autoencoders for dimensionality reduction}
  \label{fig:future_preds}
\end{figure}

Predictions that involve time-horizons beyond $(t+1)$ are typically more useful for most practical applications and can greatly increase the difficulty of the time-series prediction problem. To analyze the updated QRN's performance on different time-horizons, a new regression problem was formulated that shifted the horizons from $t = 1$ to $t = 20$. This is shown in Figure \ref{fig:future_preds}. We find that even with the added regularization that helps prevent the 256 node classical ESN from overfitting, the QRN outperforms as the time-horizon is extended, indicating better generalization and forecasting capabilities, similar to the previous section. In the naive unregularized case, it is apparent that the QRN is much better at generalizing than the larger 256 node ESN. This highlights an interesting finding in classical ML vs. QML, where it appears that very large classical models need more fine-tuning to prevent overfitting to training data and short time-horizons. 

However, the authors do note that the linear regression model performs best with larger time horizons, highlighting an important confounding factor that may arise when using classical autoencoders for dimensionality reduction. It appears that some linearization of the nonlinear dynamics associated with the chaotic KS system occurred, thus making the problem much simpler than before the autoencoder was used. Works such as \cite{Lusch2018} and \cite{azencot2020forecastingsequentialdatausing} have investigated this phenomenon and attempted to amplify it for the purpose of creating Koopman operators using autoencoders through deliberate modifications; however, in this case, the apparent linearization of our system, despite not explicitly intending it, can make interpreting results more difficult. In light of this, the authors present the results as is, while providing evidence that \textit{metrologically useful} state-preparation can enhance the performance of QRNs while increasing QFI on average.


\section{Conclusion}

In conclusion, we demonstrate the successful time-series prediction of the chaotic KS system using a QRN within a hybrid classical-quantum framework. The proposed QRN model utilized an expressive embedding scheme that also generated metrologically useful states for information processing. We also showed evidence of the QRNs increased generalization and enhanced predictive capabilities compared to classical methods as well as alternative QRN designs. To frame these results within the field of quantum metrology, we conducted experiments that examined the shot-scaling of various configurations, as well as QFI experiments showing the QRN with GHZ had improved QFI per parameter. Finally, we brought to light an important pitfall with using classical autoencoders: the clear near-linearization of the original PDE can make it much harder to determine which models are best in the latent space predictions. These findings illustrate the predictive capabilities that just a handful of qubits contain in comparison to certain classical systems, which are already fully established and have access to a much larger bank of resources. With future access to hypothetical fault-tolerant quantum computers that allow for scaling beyond 16 qubits, we speculate that the autoencoder could be completely removed from the pipeline, alleviating some of the difficulties that come with predicting high-dimensional systems and freeing us from the issue of linearized dynamics, which makes model selection more difficult. 

\section{Acknowledgements}
The authors would like to acknowledge Matthew Cook, Gerasimos Angelatos, and Vignesh Narayanan for helpful discussions. This work was sponsored by the Naval Surface Warfare Center, Panama City Division (NSWC-PCD) under the Naval Engineering Education Consortium (NEEC) Grant Program (award \#N00174-23-1-0006) and the NSWC-PCD Naval Innovative Science and Engineering (NISE) program. The author ELC acknowledges support from the DoW SMART (Science, Mathematics, and Research for Transformation) Scholarship.

\begin{table}[t]
    \centering
    \begin{tabular}{l c l}
        \hline
        \textbf{Parameter} & \textbf{Value} & \textbf{Description} \\
        \hline
        $N_{\mathrm{reservoir}}$ & $\{8,256\}$ & Reservoir size \\
        $\rho$ & $0.99$ & Spectral radius \\
        $s$ & $0.9$ & Sparsity \\
        $\alpha$ & $0.2$ & Leak rate \\
        $N_{\mathrm{train}}$ & $3000$ & Training points \\
        $N_{\mathrm{test}}$ & $2000$ & Testing points \\
        $N_{\mathrm{trials}}$ & $10$ & Random seeds \\
        $N_{\mathrm{washout}}$ & $20$ & Washout steps \\
        \hline
    \end{tabular}
    \caption{ESN hyperparameters. Different trials indicate different sets of randomly initialized weights.}
    \label{tab:esn_parameters}
\end{table}


\bibliography{sn-bibliography}

\balance

\clearpage
\appendix
\onecolumn
\appendix

\section{Supplementary Note: Observations \& Additional Results}\label{SN:Data}
\subsection{Sampled Observations}
\begin{figure*}[h!]
  \centering
  \includegraphics[width=\textwidth,keepaspectratio]{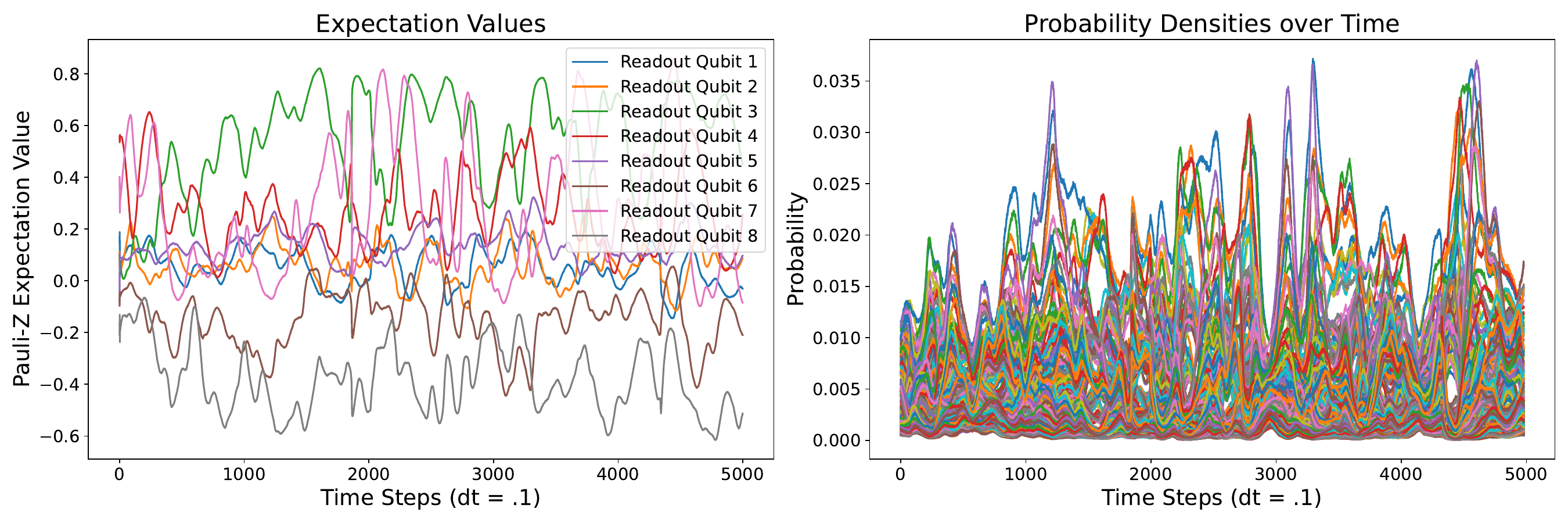}
  \caption{Plots showing both the expectation values over time and the probability densities over time of the $n_q = 16$ circuit used in \ref{subsec:pde_predictions}. These values are accumulated over $n_{shots} = 960,000$ and the probability densities are used to fit the regression model.}
  \label{fig:measurements}
\end{figure*}

Data from the QRN circuit is gathered as both probability distributions and expectation values for inspection and visualization. Below, in Figure \ref{fig:measurements}, the recorded output probability distributions for the $2^\frac{n_q}{2}$ Hilbert space and $\frac{n_q}{2}$ expectation values are shown.

\subsection{Additional Results}
\begin{figure*}[h!]
  \centering
  \includegraphics[width=\textwidth,keepaspectratio]{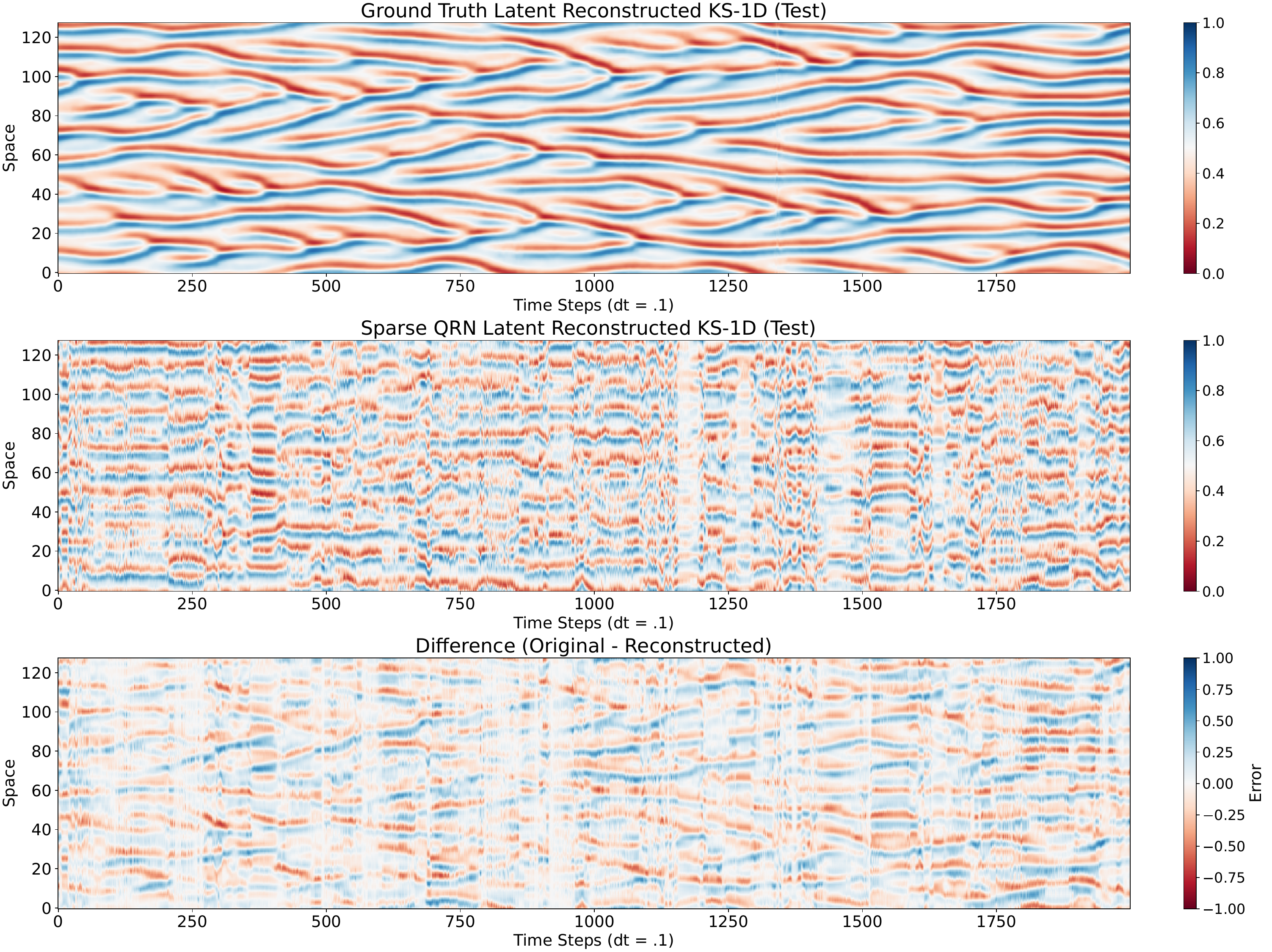}
  \caption{Plots depicting the decoded PDE prediction performance for the Sparse QRN.}
  \label{fig:sparse_results}
\end{figure*}

\begin{figure*}[h!]
  \centering
  \includegraphics[width=\textwidth,keepaspectratio]{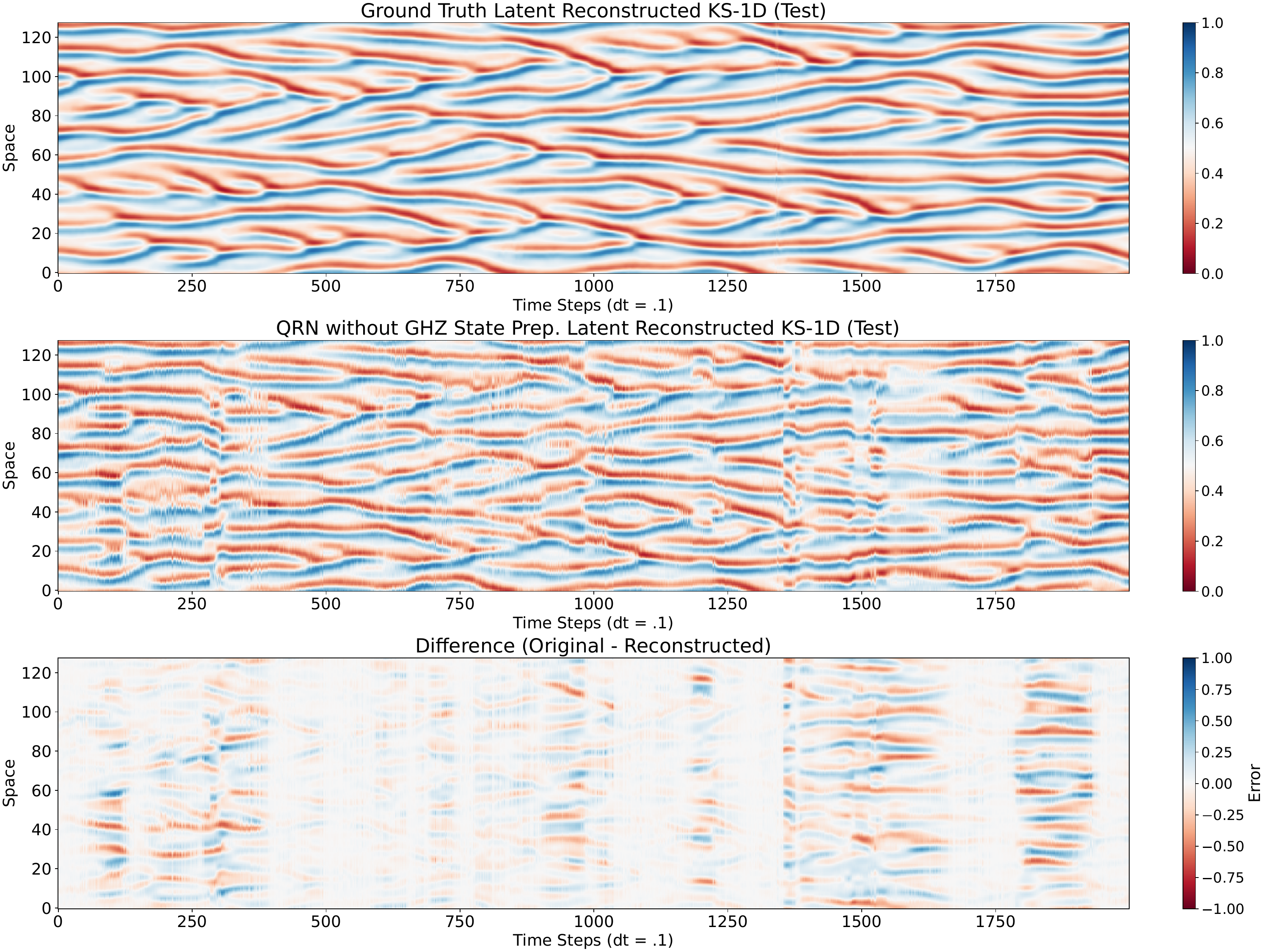}
  \caption{Plots depicting the decoded PDE prediction performance for the QRN without GHZ State Prep.}
  \label{fig:noghz_results}
\end{figure*}

\begin{figure*}[h!]
  \centering
  \includegraphics[width=\textwidth,keepaspectratio]{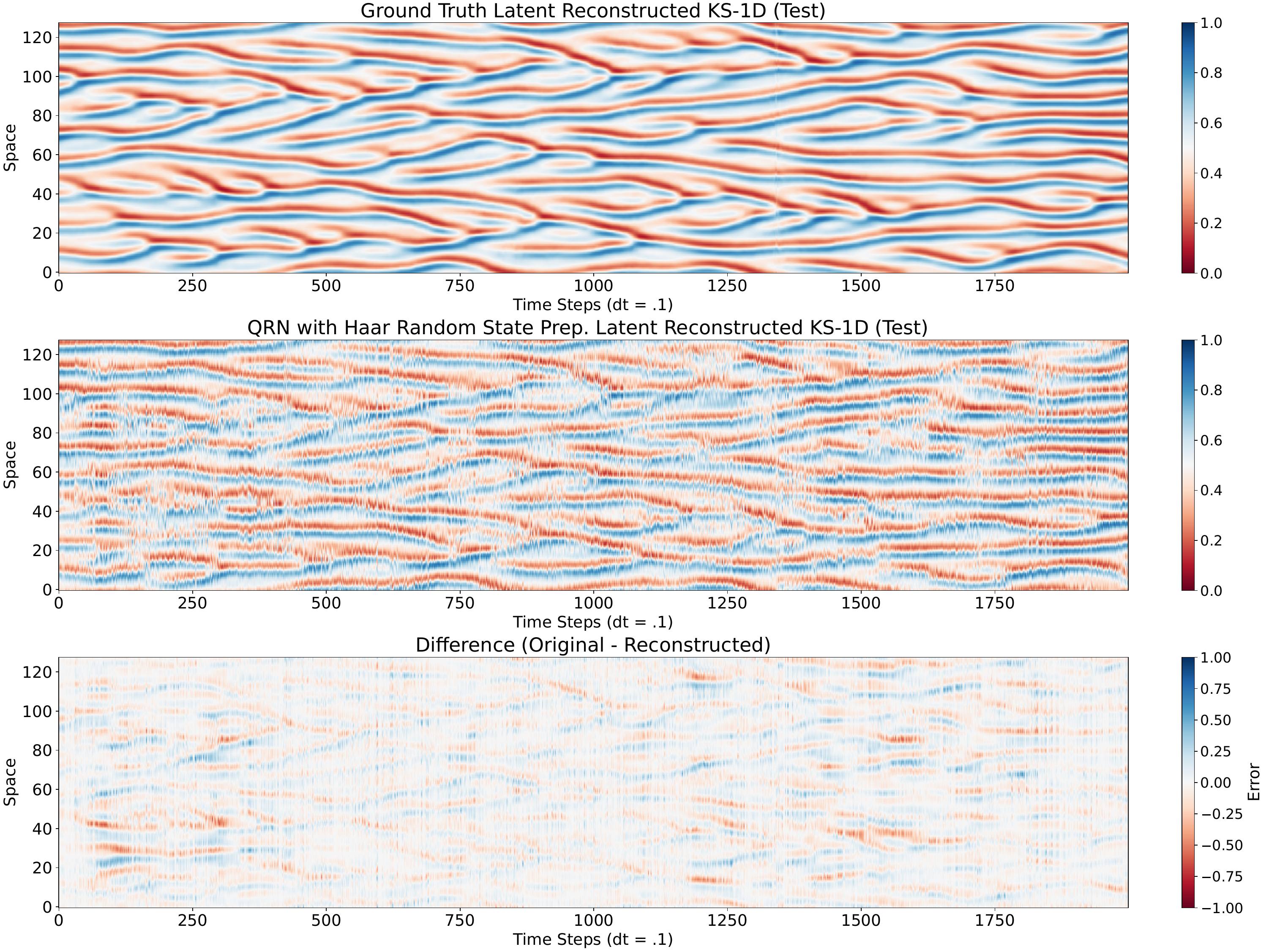}
  \caption{Plots depicting the decoded PDE prediction performance for the QRN with Haar Random State Prep.}
  \label{fig:haar_results}
\end{figure*}

The performance of the other QRN circuits shown in \ref{subsec:scaling} and Figure \ref{fig:shot_scaling_analysis} are demonstrated to be noticeably worse in the latent space RMSE comparisons, but here we also provide a look into how inaccurate the decoded outputs look with these levels of RMSE. Figure \ref{fig:sparse_results} shows the low performance of the Sparse QRN described first in \cite{Connerty2026} on the PDE time-series prediction task. For this circuit, only $n_{shots} = 10,000$ was used as this showed the best performance, as indicated in \ref{subsec:scaling}. Figure \ref{fig:noghz_results} denotes the performance of the QRN without the GHZ State Prep circuit, with $n_{shots} = 10,000$ again set for optimal performance guided by \ref{subsec:scaling}. Finally, Figure \ref{fig:haar_results} shows the performance of the QRN with Haar Random State Prep. This circuit appeared to scale with the number of shots taken, and as such $n_{shots} = 500,000$ were collected. Overall, these plots highlight the dramatic difference in performance between the differing circuits and clearly demonstrate the superiority of the QRN with GHZ State Prep depicted in Figure \ref{subsec:pde_predictions}.

\end{document}

%% file: Packages.tex
\usepackage[square, numbers]{natbib}
\usepackage{easyReview}
\usepackage[utf8]{inputenc} 
\usepackage[T1]{fontenc}    
\usepackage{hyperref}
\usepackage{url}            
\usepackage{booktabs}       
\usepackage{amsfonts}       
\usepackage{nicefrac}       
\usepackage{microtype}      
\usepackage{amsthm}
\usepackage{caption}
\usepackage{subcaption}
\usepackage{graphics} 
\usepackage{epsfig} 
\usepackage{times} 
\usepackage{amsmath} 
\usepackage{amssymb}  
\usepackage{makeidx}
\usepackage{float}
\usepackage{booktabs}
\usepackage{mathrsfs}
\usepackage{enumerate}
\usepackage{algorithm}
\usepackage{algpseudocode}
\usepackage{balance}

\usepackage{mathrsfs}  
\usepackage{xr} 
\usepackage{multicol}
\usepackage{float}
\usepackage[nolist]{acronym}
\usepackage{xcolor}

\colorlet{siaminlinkcolor}{green!50!black}
\colorlet{siamexlinkcolor}{red!50!black}

\usepackage{lipsum} 

\usepackage{cleveref} 

\usepackage[acronym]{glossaries} 

\usepackage{soul}
\usepackage{braket}
\usepackage[compat=0.4]{yquant}
\usetikzlibrary{quantikz2}

\usepackage{scalerel}

\usepackage{xstring}
\usetikzlibrary{fit,positioning}

\usetikzlibrary{decorations.pathmorphing,positioning}

%% file: figures/nn_GHZ.tex

\begin{tikzpicture}[shorten >=1pt, node distance=2cm and 3cm]
    \tikzstyle{unit}=[draw,shape=circle,minimum size=1.1cm,fill=white]

    \begin{scope}[shift={(-2.65,0)}]
      \begin{yquant*}
    
    

        [name=past2a] qubit {$\ket{\psi_{mem}}$} t0;
        [name=past2b] qubit {$\ket{0}$} t0b;

        [name=past1a] qubit {$\ket{\psi_{mem}}$} t1;
        [name=past1b] qubit {$\ket{0}$} t1b;

        [name=pasta] qubit {$\ket{\psi_{mem}}$} t2;
        [name=pastb] qubit {$\ket{0}$} t2b;
    nobit out;
    
    hspace {1mm} -;
    
    
    [this subcircuit box style={dashed, "Recurrent Block $\mathcal{B}$ $\times \; T$"}]
    subcircuit {
        [inout] 
        qubit {}t0;
        qubit {}t0b;
        qubit {}t1;
        qubit {}t1b;
        qubit {}t2;
        qubit {}t2b;
        nobit out;


    [fill=red!35] box {$GHZ$} (t0,t0b,t1,t1b,t2,t2b); 
    
    [this subcircuit box style={dashed, "Reuploading Block $\times \; n_c$"}]
    subcircuit {
        [inout] 
        qubit {}t0;
        qubit {}t0b;
        qubit {}t1;
        qubit {}t1b;
        qubit {}t2;
        qubit {}t2b;
    
    [fill=green!35] box {$R(\vec\Theta^{(t)}_{1,1})$} t0; 
    [fill=green!35] box {$R(\vec\Theta^{(t)}_{2,1})$} t0b;
    cnot t0b | t0;
    [fill=green!35] box {$R(\vec\Theta^{(t)}_{1,2})$} t0;
    [fill=green!35] box {$R(\vec\Theta^{(t)}_{2,2})$} t0b;
    [fill=orange!35] box {$R_Y(\theta^{(t)}_{1})$} t0b | t0;
    [fill=green!35] box {$R(\vec\Theta^{(t)}_{1,3})$} t0;
    [fill=green!35] box {$R(\vec\Theta^{(t)}_{2,3})$} t0b;
    [fill=yellow!35] box {$R_X(\theta^{(t)}_{4})$} t0b | t0;
    
    [fill=green!35] box {$R(\vec\Theta^{(t)}_{3,1})$} t1;
    [fill=green!35] box {$R(\vec\Theta^{(t)}_{4,1})$} t1b;
    cnot t1b | t1;
    [fill=green!35] box {$R(\vec\Theta^{(t)}_{3,2})$} t1;
    [fill=green!35] box {$R(\vec\Theta^{(t)}_{4,2})$} t1b;
    [fill=orange!35] box {$R_Y(\theta^{(t)}_{2})$} t1b | t1;
    [fill=green!35] box {$R(\vec\Theta^{(t)}_{3,3})$} t1;
    [fill=green!35] box {$R(\vec\Theta^{(t)}_{4,3})$} t1b;
    [fill=yellow!35] box {$R_X(\theta^{(t)}_{5})$} t1b | t1;
    
    [fill=green!35] box {$R(\vec\Theta^{(t)}_{5,1})$} t2;
    [fill=green!35] box {$R(\vec\Theta^{(t)}_{6,1})$} t2b;
    cnot t2b | t2;
    [fill=green!35] box {$R(\vec\Theta^{(t)}_{5,2})$} t2;
    [fill=green!35] box {$R(\vec\Theta^{(t)}_{6,2})$} t2b;
    [fill=orange!35] box {$R_Y(\theta^{(t)}_{3})$} t2b | t2;
    [fill=green!35] box {$R(\vec\Theta^{(t)}_{5,3})$} t2;
    [fill=green!35] box {$R(\vec\Theta^{(t)}_{6,3})$} t2b;
    [fill=yellow!35] box {$R_X(\theta^{(t)}_{6})$} t2b | t2;
    
    [fill=blue!35] box {$CRZ(\vec\theta^{(t)})$} (t0,t0b,t1,t1b,t2,t2b); 
    
    } (t0,t0b,t1,t1b,t2,t2b);

    
    measure t0b;
    measure t1b; 
    measure t2b;
    
    text {$M^1_0$} out | t0b;
    text {$M^2_0$} out | t1b;
    text {$M^3_0$} out | t2b;
    
    discard t0b,t1b,t2b;
    box {$\ket{0}$} t0b;
    box {$\ket{0}$} t1b;
    box {$\ket{0}$} t2b;
    settype {qubit} t0b;
    settype {qubit} t1b;
    settype {qubit} t2b;
    
    } (t0,t0b,t1,t1b,t2,t2b,out);

    \end{yquant*}
    \end{scope}

\end{tikzpicture}

%% file: figures/CRZ.tex
\small
\tikzset{
  yquant/every qubit/.style={minimum height=2ex},
  yquant/every box/.style={inner sep=0.3mm},
}

\begin{tikzpicture}

      \begin{yquant*}

        [name=past2a] qubit {$\ket{\psi_{mem}}$} t0;
        [name=past2b] qubit {$\ket{\psi_{out}}$} t0b;

        [name=past1a] qubit {$\ket{\psi_{mem}}$} t1;
        [name=past1b] qubit {$\ket{\psi_{out}}$} t1b;

        [name=pasta] qubit {$\ket{\psi_{mem}}$} t2;
        [name=pastb] qubit {$\ket{\psi_{out}}$} t2b;
    nobit out;
    
    hspace {1mm} -;

    [this subcircuit box style={dashed, "CRZ"}]
    subcircuit {
        [inout] 
        qubit {}t0;
        qubit {}t0b;
        qubit {}t1;
        qubit {}t1b;
        qubit {}t2;
        qubit {}t2b;
        nobit out;

    [fill=blue!35] box {$R_Z(\theta_{7})$} t0b | t0;
    [fill=blue!35] box {$R_Z(\theta_{8})$} t1 | t0b;
    [fill=blue!35] box {$R_Z(\theta_{9})$} t1b | t1;
    [fill=blue!35] box {$R_Z(\theta_{10})$} t2 | t1b;
    [fill=blue!35] box {$R_Z(\theta_{11})$} t2b | t2;

    } (t0,t0b,t1,t1b,t2,t2b,out);

    \end{yquant*}

\end{tikzpicture}

%% file: figures/ghz.tex
\begin{tikzpicture}[]

      \begin{yquant*}

        [name=past2a] qubit  {$\ket{\psi_{mem}}$} t0;
        [name=past2b] qubit {$\ket{0}$} t0b;

        [name=past1a] qubit  {$\ket{\psi_{mem}}$} t1;
        [name=past1b] qubit  {$\ket{0}$} t1b;

        [name=pasta] qubit  {$\ket{\psi_{mem}}$} t2;
        [name=pastb] qubit {$\ket{0}$} t2b;
    nobit out;
    
    hspace {1mm} -;

    [this subcircuit box style={dashed, "GHZ"}]
    subcircuit {
        [inout] 
        qubit {}t0;
        qubit {}t0b;
        qubit {}t1;
        qubit {}t1b;
        qubit {}t2;
        qubit {}t2b;
        nobit out;

    h t1;
    
    cnot t1b | t1;
    cnot t0b | t1;
    cnot t2 | t1b;
    cnot t2b | t2;
    cnot t0 | t0b;

    } (t0,t0b,t1,t1b,t2,t2b,out);

    \end{yquant*}

\end{tikzpicture}